\documentclass[letterpaper, 10 pt, conference]{ieeeconf}  

\IEEEoverridecommandlockouts                              

\usepackage{graphics} 
\usepackage{epsfig} 
\usepackage{mathptmx} 
\usepackage{times} 
\usepackage{amsmath} 
\usepackage{amssymb}  
\usepackage{multirow}
\usepackage[hidelinks]{hyperref}
\usepackage[table,xcdraw]{xcolor}
\usepackage{colortbl}
\usepackage{subcaption}
\usepackage{xcolor}

\title{\LARGE \bf Towards simultaneous decoding of kinetic and kinematic movement parameters during grasp and lift task by non-invasive brain imaging
}

 \author{ Parth G. Dangi$^{1}$, Yogesh Kumar Meena$^{1}$
 \thanks{$^{1}$ Parth G. Dangi and Yogesh Kumar Meena are with Human-AI Interaction (HAIx) Lab, IIT Gandhinagar, India
          {\tt\small yk.meena@iitgn.ac.in}}%
  }

\begin{document}

\maketitle

\begin{abstract}

Brain–machine interfaces (BMIs) can assist individuals with limited mobility, such as stroke survivors or amputees. One of the key challenges in developing BMIs is expanding their usability and control, which can be achieved by accurately decoding multiple kinematic and kinetic parameters. To address this, we propose three regression models—partial least squares regressor, multi-layered perceptron, and attention-based regressor—to decode multiple movement parameters from EEG signals. We evaluated these models on the WAY-EEG-GAL dataset, focusing on their performance under subject-specific and subject-independent conditions with two strategies: a single model for all parameters and a baseline with separate models for each parameter. Among all regressors, the attention-based regressor achieved the best performance, with an $R^2$ of 0.8 and a latency of 29.2 milliseconds, demonstrating significant improvement in simultaneous multi-parameter decoding. However, its performance dropped for single-parameter decoding. The multi-layered perceptron showed more consistent but lower accuracy across both decoding types ($R^2$ = 0.49). These findings highlight the potential of attention-based models for real-time multi-command BMI systems and contribute to the development of more intuitive control devices.

\end{abstract}






\section{Introduction}

A brain-machine interface (BMI) translates brain activity into commands that can control hardware devices~\cite{1}. This technology has the unique ability to bypass muscles and peripheral nerves, allowing it to assess movement intentions directly. It has been used in developing effective stroke rehabilitation tools and prosthetic devices that provide engaging and intuitive control~\cite{3}. Additionally, there is ongoing exploration of these devices for commercial use, aiming to provide a more intuitive and efficient way to control various technologies~\cite{1, 3}. However, the current BMI devices rely on classifiers~\cite{1}, which limits the number of commands they can generate~\cite{meena2015simultaneous}. While many studies have attempted to improve classifiers by incorporating finer movements~\cite{6} and by classifying multiple movement parameters, such as speed, direction, and force, simultaneously~\cite{7}, the finite number of commands generated by these BMIs limits their control capabilities~\cite{3, 7318410}. This limitation reduces usability, particularly for individuals with restricted mobility due to conditions such as stroke or amputation~\cite{8}.



To address these limitations, non-invasive regressor-based BMIs have been proposed to increase the number of commands and, more importantly, to provide greater control over devices~\cite{9}. While initial studies were limited by the lack of robust regressor models and methods of preprocessing the non-invasive brain activity signal like electroencephalography (EEG)~\cite{1, 3}, recent models have shown success in decoding specific parameters, such as force~\cite{9, 11} and movement~\cite{12}, with high precision across trajectories, improving the user's control over the system~\cite{8}. Some studies have created models to decode multiple kinematic parameters, like position and velocity, from EEG data~\cite{27}.


While these models have shown great precision in decoding specific parameters, their inability to predict multiple movement parameters of varying nature (e.g., linear position and angular rotation, or linear position and force exerted) during a movement has limited their performance~\cite {9, 11, 12}. Furthermore, the specific parameter decoded by these regressors cannot be visualised properly without getting feedback from other parameters~\cite{13}. Hence, these regressor-based models must decode multiple parameters simultaneously to accurately reproduce movements, thereby increasing control over devices and the usability of BMI systems~\cite{8}.

To achieve this end, most studies have used invasive methods to observe brain activity in human and non-human subjects~\cite{14}. However, similar studies are not performed using non-invasive methods such as EEG due to poor signal quality and low spatial resolution~\cite{3}. Prior studies addressed these problems by developing advanced regressor models~\cite{9} or preprocessing and feature-extraction pipelines~\cite{11} that isolate the EEG signal from specific regions and predict specific movement parameters~\cite{9, 11}. While these models have shown high accuracy in decoding single parameters, they and related methods have not been applied to decode multiple movement parameters. 

In this work, we address this limitation by proposing a framework to simultaneously decode multiple kinetic and kinematic movement parameters from non-invasive EEG signals. This framework is developed by utilising the EEGForceMap pipeline~\cite{11} to decode multiple parameters during a grasp-and-lift task, and integrating it with three regressor models. These regressor models represent traditional, deep-learning-based, and attention-based regression approaches. Furthermore, we compare these regressors in terms of precision and feasibility for real-world deployment. The main contributions of this work are:

\begin{figure*}[ht!] 
\centering
\includegraphics[width=1.0\linewidth]{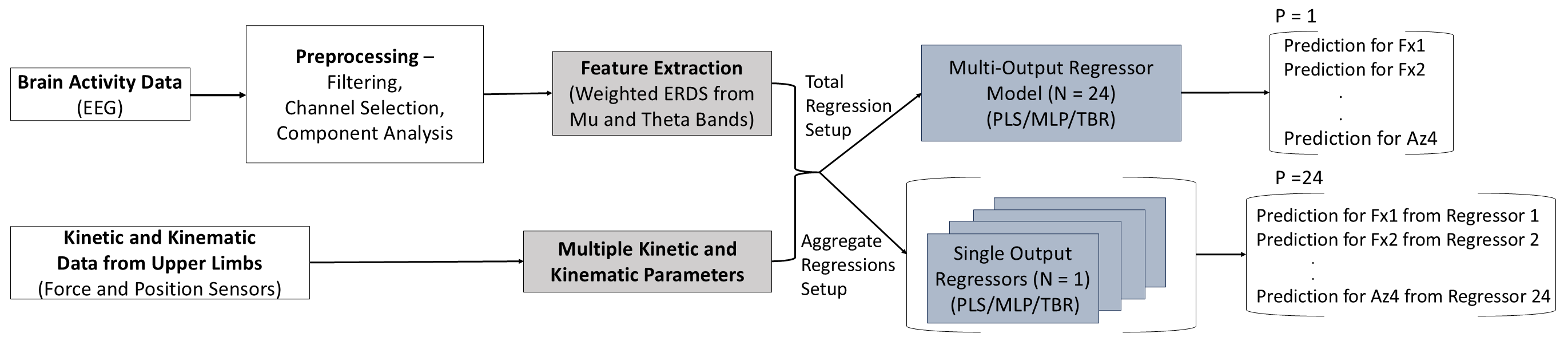}
\caption{The proposed framework describes a method for preprocessing EEG signals and extracting features with the EEGForceMap approach~\cite{11}. It employs a regression setup to decode multiple parameters, where \( N \) denotes the outputs from individual regressor models and \( P \) represents the total number of outputs. The regression techniques used include partial least squares (PLS), multi-layer perceptron (MLP), and transformer-based regressor (TBR).}


\label{Figure 1}
\end{figure*} 

\begin{itemize}
    \item Proposing the attention-based regression approach in decoding the EEG signal by developing a transformer-based regressor.
    \item Developing a partial least squares regressor and a multi-layered perceptron-based regressor alongside the transformer-based regressor to decode both kinematic and kinetic movement parameters continuously, and evaluating their performance. 
    \item Scaling up the ERDS feature set from the EEGForceMap approach~\cite{11} to analyse the representation of movement parameters and identify patterns in their neural encoding for simultaneous decoding.

\end{itemize}

\section{METHODS AND MATERIALS}
\label{Section 2}

To decode multiple movement parameters simultaneously, we developed a transformer-based regressor with an architecture similar to the attention-based classifier used in prior studies~\cite{21}. We compared it with two widely used regressor models~\cite{9, 11, 24}. All these regressor models were integrated with the EEGForceMap approach~\cite{11}, and tested in pseudo-online environments. The methodology for achieving these steps is detailed in the following section. 


\subsection{Dataset}

For evaluating the regressor models, grasp-and-lift trial data from the WAY-EEG-GAL dataset were used. The WAY-EEG-GAL dataset, developed by Luciw et al.~\cite{17}, records brain activity during grasp and lift trials across different weight and surface conditions from 12 participants~\cite{17}. In this work, we focused on trials with all participants. We varied only the weight while keeping the surface constant. This approach allowed for a clear presentation of multiple movement parameters. From this dataset, the EEG and kinematic parameters were extracted. The parameters targeted were divided into three domains: \textit{Kinetic} (the force exerted by the fingers), \textit{Linear Kinematic} (the position of the fingers and wrist in 3D space) and \textit{Angular Kinematic} (rotation of the fingers and the wrist around coordinate axes in 3D space). Twenty-four movement parameters were selected, as listed in Table~\ref{Table 1}. The prepared EEG data, labelled with these 24 movement parameters, were later preprocessed for feature extraction and regression.

\subsection{Preprocessing and Feature Extraction}

The regressor models were integrated into the EEGForceMap approach~\cite{11}, which leverages cognitive science principles to extract features suitable for lightweight, real-time regressors. Although EEGForceMap utilised three feature sets, Event-Related Desynchronisation/Synchronisation (ERDS) demonstrated the strongest correlation with movement parameters and was described as the most suitable feature set for regression~\cite{11}. Consequently, we extracted only ERDS features from nine channels covering the premotor-parietal regions~\cite{11}, specifically up-weighting the Mu (9-11~Hz) and Theta (4-7~Hz) bands. This emphasis highlights premotor-parietal activity, where movement encoding is most prominent during planning and imagery~\cite{18}. Figure~\ref{Figure 1} illustrates the implementation of the integrated EEGForceMap approach~\cite{11} and the corresponding regressor architectures.

\begin{table*}[ht!]
\centering
\caption{List of parameters from WAY-EEG-GAL dataset selected for simultaneous decoding (GAL - Grasp and Lift)}
\label{Table 1}
\resizebox{0.9\textwidth}{!}{%
\begin{tabular}{l|l|l|l}
\hline
\textbf{Parameter} &
  \textbf{Sampling Location} &
  \textbf{Parameter Description} &
  \textbf{Domain Description} \\ \hline
Fx1 &
  Index Finger &
  Force Exerted by index finger on an object along the X-axis &
  \multirow{6}{*}{\begin{tabular}[c]{@{}l@{}}Force exerted by thumb \\ and index finger on the \\ object across X, Y and \\ Z axis during the GAL \\ task\end{tabular}} \\
Fy1 &
  Index Finger &
  Force Exerted by index finger on an object along the Y-axis &
   \\
Fz1 &
  Index Finger &
  Force Exerted by index finger on an object along the Z-axis &
   \\
Fx2 &
  Thumb &
  Force Exerted by thumb on an object along the X-axis &
   \\
Fy2 &
  Thumb &
  Force Exerted by thumb on an object along the Y-axis &
   \\
Fz2 &
  Thumb &
  Force Exerted by thumb on an object along the Z-axis &
   \\ \hline
Px2 &
  Index Finger &
  Position of the index finger during the movement (X-coordinate) &
  \multirow{9}{*}{\begin{tabular}[c]{@{}l@{}}Position of thumb, index \\ finger and wrist during \\ the GAL task in a 3D \\ space as described in \\ (X, Y, Z) coordinates\end{tabular}} \\
Py2 &
  Index Finger &
  Position of the index finger during the movement (Y-coordinate) &
   \\
Pz2 &
  Index Finger &
  Position of the index finger during the movement (Z-coordinate) &
   \\
Px3 &
  Thumb &
  Position of the thumb during the movement (X-coordinate) &
   \\
Py3 &
  Thumb &
  Position of the thumb during the movement (Y-coordinate) &
   \\
Pz3 &
  Thumb &
  Position of the thumb during the movement (Z-coordinate) &
   \\
Px4 &
  Wrist &
  Position of the wrist during the movement (X-coordinate) &
   \\
Py4 &
  Wrist &
  Position of the wrist during the movement (Y-coordinate) &
   \\
Pz4 &
  Wrist &
  Position of the wrist during the movement (Z-coordinate) &
   \\ \hline
Ae2 &
  Index Finger &
  Angular displacement of index finger along X-axis &
  \multirow{9}{*}{\begin{tabular}[c]{@{}l@{}}Angular Rotation seen \\ in thumb and index \\ finger and wrist along \\ the X-axis (elevation (e)), \\ Y-axis (roll (r)), and \\ Z-axis (azimuth (z)) \\ in the GAL task\end{tabular}} \\
Ar2 &
  Index Finger &
  Angular displacement of index finger along the Y-axis &
   \\
Az2 &
  Index Finger &
  Angular displacement of index finger along the Z-axis &
   \\
Ae3 &
  Thumb &
  Angular displacement of thumb along the X-axis &
   \\
Ar3 &
  Thumb &
  Angular displacement of thumb along the Y-axis &
   \\
Az3 &
  Thumb &
  Angular displacement of thumb along the Z-axis &
   \\
Ae4 &
  Wrist &
  Angular displacement of wrist along the X-axis &
   \\
Ar4 &
  Wrist &
  Angular displacement of wrist along the Y-axis &
   \\
Az4 &
  Wrist &
  Angular displacement of wrist along the Z-axis &
   \\ \hline
\end{tabular}%
}
\end{table*}

\subsection{Feature Space Analysis}

Before evaluating the regressor models, the ERDS features extracted by the EEGForceMap approach were analysed to examine the encoding of movement parameters across frequency bands. First, the Pearson correlation between each parameter and its ERDS representation was computed from the Mu (9-11 Hz) and Theta (4-7 Hz) band signals that originated from the premotor-parietal regions. The ERDS values and the Pearson correlations across the Mu and Theta frequency bands were compared using Student's t-test. Comparisons of ERDS representations and Pearson correlations were conducted for within-group (kinetic-kinetic or kinematic-kinematic feature pairs) and across-group (kinetic-kinematic) pairs, with each movement parameter compared with the others to identify closely similar parameters. 


\subsection{Regressor Models}


To decode multiple movement parameters, we implemented three regressor models, as shown in Figure~\ref{Figure 1}, and evaluated their performance.



\subsubsection{Partial Least Squares Regressor} The partial least squares regressor (PLS) is commonly used in the state-of-the-art work for decoding single or multiple movement parameters within the same domain, outperforming other traditional regressors like linear regressors~\cite{11, 24}. For this reason, we utilised the PLS model, optimising it through adaptive hyper-parameter tuning and selecting the number of principal components based on training-phase accuracy. We also minimised the mean squared error (MSE) during training to enhance the model's performance.


\subsubsection{Multi-layered Perceptron}
Traditional regressors like PLS, are complemented by deep learning-based regressors in BMI systems for efficient neural signal decoding~\cite{9, 19}.  Among these methods, the multi-layer perceptron (MLP) is considered a simpler alternative to the more complex deep learning architectures used in BMI systems~\cite{28}. In this study, the MLP had three layers with 512, 256, and 128 nodes, and its performance was enhanced using the Adam optimiser.


%

\subsubsection{Transformer-based Regressor} 

Along with traditional and deep-learning-based regression approaches, an attention-based decoding approach was also used with a transformer-based regressor (TBR), which has seen increased use in BMI applications in recent years due to its architecture, which enables the identification of patterns and correlations across time steps~\cite{21}. Despite their potential for regression, the attention-based decoding approach has been centred on classifiers that generate only a limited number of commands~\cite{21}. Hence, TBR has been developed to test its validity for decoding multiple movement parameters simultaneously. The regressor's architecture consisted of a 4-layer encoder with 32 model dimensions and 4 attention heads. The model leverages a 256-dimensional feed-forward network to capture non-linear feature interactions.

\subsection{Experiment Setup}

To evaluate the feasibility of simultaneously decoding multiple movement parameters, we trained regressor models under both subject-specific and subject-independent conditions. The subject-specific models were developed and tested using EEG data from individual participants selected from the WAY-EEG-GAL dataset~\cite{17}. In contrast, the subject-independent models employed the leave-one-subject-out (LOSO) method, in which the model was trained on EEG data from all participants except one, with the excluded participant's data used for testing. These regressor models were designed to continuously decode all movement parameters simultaneously under both subject-specific and subject-independent conditions.



In parallel with the simultaneous decoding models, a baseline configuration was developed using a similar set of architectures for each regressor type. This design was structured so that each regressor independently predicted a single movement parameter, mirroring the methodology used in the EEGForceMap implementation in a prior study~\cite{11}. This approach produced a set of models that yielded aggregate predictions from 24 regressors, each designed to predict a specific EEG signal parameter. The outputs of all regressors were combined to provide a full profile of the movement task in both subject-specific and subject-independent conditions. This modular arrangement of decoders, which served as the baseline, was called \textit{aggregate} decoders. In contrast, the single-decoder model that decoded all parameters simultaneously was referred to as a \textit{total} decoder. Descriptions of both regressors are illustrated in Figure~\ref{Figure 1}.

These regressors were trained under pseudo-online conditions using a 0.1-second sliding window with a 0.05-second step size. The data was split into training, testing, and validation sets in a 7:2:1 ratio. For TBR, the sequence length for regressor training was set to 50 tokens to ensure the transformer could effectively capture patterns from relatively small EEG datasets. To prevent overfitting, a 30$\%$ dropout of input features is applied during training for MLP and TBR regressors. Each regressor model was trained for 100 epochs. 

\subsection{Performance metrics}


The regressors were assessed for both their ability to decode multiple movement parameters simultaneously and their feasibility for deployment in a closed-loop BMI setup. To evaluate both of these aspects, three metrics were used, which are described below:

\begin{itemize}
    \item Coefficient of Determination (CoD): To check for accuracy, i.e. similarity of the prediction from a regressor with the true value given in the WAY-EEG-GAL dataset, the coefficient of determination (also known as $R^2$ value) was calculated post-hoc.

    \item Mean squared error (MSE): While the accuracy has been calculated, the precision of these predictions across different participants was also calculated by calculating the mean squared error across participants.
    
    \item Latency: While decoding performance was calculated, the feasibility of real-world deployment was also evaluated. For this, the latency, the time taken to preprocess, feature extraction and prediction, was calculated 
\end{itemize}

\begin{figure}[t!]
     \begin{subfigure}[]{0.43\textwidth}
         \centering
         \includegraphics[trim={0cm 0cm 0cm 0cm},width=\textwidth]{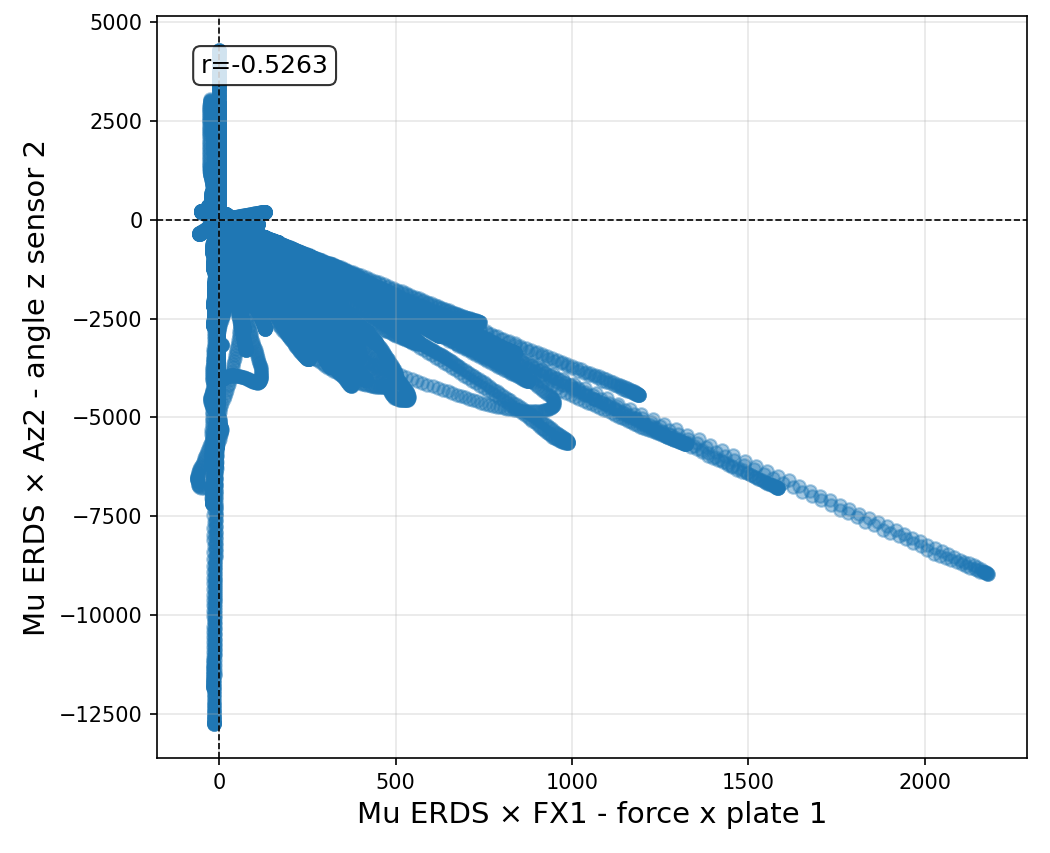}
     \end{subfigure}
     \begin{subfigure}[]{0.43\textwidth}
         \centering
         \includegraphics[trim={0cm 0cm 0cm 0cm},width=\textwidth]{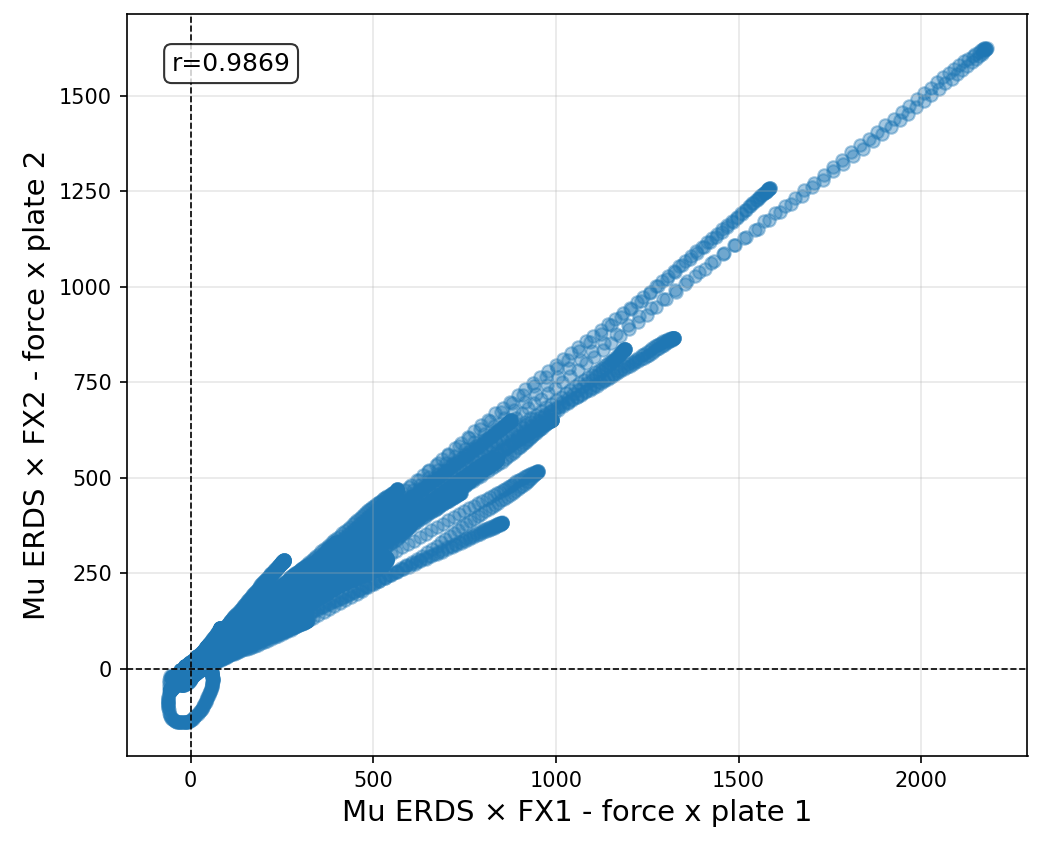}
     \end{subfigure}
    \caption{A sample of scatter plots showing autocorrelation across different movement parameters' ERDS activity in $\mu$ band for between-group (top), and within-group (bottom) parameters (Az2: rotation of index finger along Z-axis; Fx1: force applied by the thumb along the X-axis; Fx2: force applied by the index finger along the X-axis)}
    \label{Figure 3}
\end{figure}


All of these metrics were averaged across all parameters and later analysed for significance using two-way ANOVA, with the regressor models used for decoding and the strategies of their implementation as co-factors. For the post hoc analysis, Tukey's HSD test was performed to identify differences in regressor and strategy type that were significant relative to others. Finally, an Eta squared effect size was calculated to identify the effect of the difference on the performance of the regressor models.

\section{Result}
\label{Section 3}

\subsection{Comparison of Features across Parameters}


\begin{table*}[]
\centering
\caption{Performance comparison of each regressor using different regression methods under subject-specific (SS) and subject-independent (SI) conditions. Metrics presented include Coefficient of Determination (CoD) and Mean Squared Error (MSE)}
\label{Result Table}
\resizebox{0.9\textwidth}{!}{%
\begin{tabular}{l|llllll|lccccc}
\hline
 &
  \multicolumn{6}{l|}{\textbf{Aggregate Regression Setup (Baseline)}} &
  \multicolumn{6}{l}{\textbf{Total Regression Setup}} \\ \cline{2-13} 
 &
  \multicolumn{2}{l|}{\textbf{CoD}} &
  \multicolumn{2}{l|}{\begin{tabular}[c]{@{}l@{}} \textbf{MSE}\end{tabular}} &
  \multicolumn{2}{l|}{\begin{tabular}[c]{@{}l@{}}\textbf{Latency (ms)}\end{tabular}} &
  \multicolumn{2}{l|}{\textbf{CoD}} &
  \multicolumn{2}{l|}{\begin{tabular}[c]{@{}l@{}} \textbf{MSE}\end{tabular}} &
  \multicolumn{2}{l}{\begin{tabular}[c]{@{}l@{}}\textbf{Latency (ms)}\end{tabular}} \\ \cline{2-13} 
\multirow{-3}{*}{\textbf{Regressor}} &
  \textbf{SS} &
  \multicolumn{1}{l|}{\textbf{SI}} &
  \textbf{SS} &
  \multicolumn{1}{l|}{\textbf{SI}} &
  \textbf{SS} &
  \textbf{SI} &
  \textbf{SS} &
  \multicolumn{1}{l|}{\textbf{SI}} &
  \multicolumn{1}{l}{\textbf{SS}} &
  \multicolumn{1}{l|}{\textbf{SI}} &
  \multicolumn{1}{l}{\textbf{SS}} &
  \multicolumn{1}{l}{\textbf{SI}} \\ \hline
PLS &
  0.03 &
  \multicolumn{1}{l|}{0.08} &
  41.52 &
  \multicolumn{1}{l|}{62.32} &
  266.07 &
  266.17 &
  \multicolumn{1}{c}{0.13} &
  \multicolumn{1}{c|}{0.06} &
  \cellcolor[HTML]{FFFFFF}41.52 &
  \multicolumn{1}{c|}{62.62} &
  265.38 &
  265.41 \\
MLP &
  \textbf{0.31} &
  \multicolumn{1}{l|}{\textbf{0.2}} &
  \textbf{19.55} &
  \multicolumn{1}{l|}{\textbf{48.05}} &
  267.32 &
  265.83 &
  \multicolumn{1}{c}{0.49} &
  \multicolumn{1}{c|}{0.24} &
  19.56 &
  \multicolumn{1}{c|}{45.56} &
  265.63 &
  266.71 \\
TBR &
  0.01 &
  \multicolumn{1}{l|}{0.02} &
  43.67 &
  \multicolumn{1}{l|}{44.01} &
  \textbf{60.09} &
  \textbf{66.39} &
  \multicolumn{1}{c}{\textbf{0.80}} &
  \multicolumn{1}{c|}{\textbf{0.34}} &
  \textbf{4.55} &
  \multicolumn{1}{c|}{\textbf{21.53}} &
  \textbf{30.34} &
  \textbf{46.41} \\ \hline
\end{tabular}%
}
\end{table*}

To evaluate the representation of kinetic and kinematic aspects within the ERDS feature set, we conducted a feature-space analysis, showing moderate correlation between the ERDS features and the parameters they represent across the Mu (Pearson's r: 0.25) and Theta (Pearson's r: 0.23) bands. Furthermore, no significant difference between the bands was seen for correlations (t-test: 0.691, p-value = 0.49). When the correlation between ERDS representations within each group was calculated, the ERDS representations, as shown in Figure~\ref{Figure 3}, demonstrated a high correlation in parameter values (Pearson's r: 0.89) across bands. Through further analysis, we observed that the within-group correlation (Pearson's r: 0.96) was significantly higher than the between-group correlation (Pearson's r: 0.82) (t-score: 3.25, p-value = 0.002) with a lower effect size (Cohen's d: 0.25).

\subsection{Comparison of Regressor Performance}


After analysing the neural representations of the movement parameters in the feature set, the performance of the regressor models was evaluated using the data analysis methods discussed in the prior section. During the evaluation of regressor models' performance, we observed that the metrics for each regressor and strategy are shown to be significantly different ($f$-value: 666.37, $p$-value $<$ 0.0001), with a large effect size ($\eta$-value = 0.865), with no significant difference in the matrices across participants ($f$-value: 0.21, $p$-value: 0.795). The performance results of all the regressor models, strategies and conditions are recorded in Table~\ref{Result Table}



\subsubsection{Partial Least Squares Regressor}

\begin{figure}[ht!] 
\includegraphics[width=0.5\textwidth]{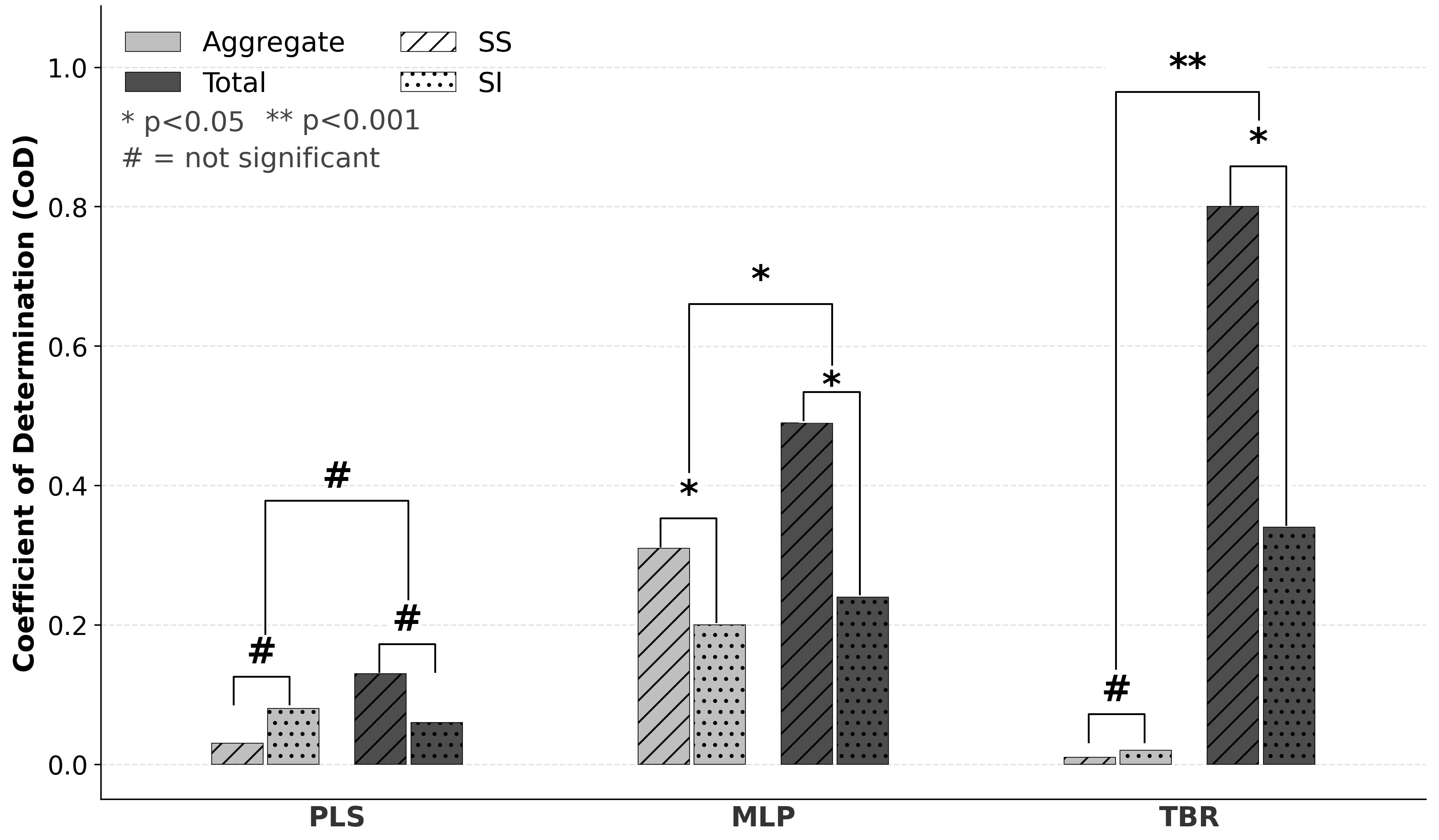}
\caption{Bar plot showing the accuracy of partial least square (PLS) regressor, multi-layered perceptron (MLP) regressor, and transformer-based regressor (TBR) under total and aggregate strategies in subject-specific (SS) and subject-independent (SI) settings}
\label{CoD Results}
\end{figure}

The PLS regressor is a traditional model that has been used repeatedly to predict movement parameters~\cite{24}. While successful in prior studies~\cite{24}, it achieved significantly lower accuracy ($0.13 \pm 0.03$) when decoding all parameters simultaneously, and while decoding single parameters ($0.05 \pm 0.02$), as shown in Figure~\ref{CoD Results}. When precision was checked, it showed the highest mean square error ($41.52 \pm 10.14$). Finally, the models' latency was also higher ($265.38 \pm 45.23 $ ms), indicating that PLS regressor models are not feasible for real-world deployment. This trend was observed across both total and aggregate regressor models, indicating that the traditional regressor models used in prior MI studies~\cite{24} show significantly poor performance (F-score = 1945.75, p-value $<$ 0.0001) at decoding multiple movement parameters simultaneously. 


\subsubsection{Multi-layered Perceptron}

Compared with the PLS regressor, the MLP, a deep learning-based regressor, achieves significantly higher accuracy ($0.49 \pm 0.03$) for decoding multiple movement parameters simultaneously, as shown in Figure~\ref{CoD Results}. With significantly lower mean square error ($19.41 \pm 6.23$) than that of PLS. This performance was consistent across both total and aggregate strategies (T-score = 1.24, F-value $>$ 0.05). Furthermore, when latency was calculated for MLP, it was similar to that of PLS in the total regression strategy ($265.38 \pm 36.31$ ms), indicating a significant improvement over traditional regression models (F-value = 1514.36, p-value $<$ 0.0001). However, the accuracy of each parameter varied significantly (F-value = 2.76, p-value $<$ 0.05). Indicating towards the inability of the MLP regressor to decode each parameter separately, when all of them had to be decoded from the same data.

\subsubsection{Transformer-based Regressor}

The newly proposed TBR, which utilises an attention-based regression approach, has shown better performance in decoding multiple movement parameters simultaneously, significantly achieving high accuracy ($0.8 \pm 0.09$) across all kinetic and kinematic parameters for a grasp-and-lift task (F-score = 1428.21, p-value $<$ 0.0001), as shown in Figure~\ref{CoD Results} with least mean square error ( $4.54 \pm 1.28$). and latency (Latency = $30.33 \pm 9.32$ ms), indicating its feasibility for real-world deployment. However, these changes were only seen in the total regression strategy. When transformer-based regression was implemented for the aggregate strategy, the TBR's accuracy (0.01) was lower with high mean square error (43.94), indicating a significant drop in performance (F-value = 234.61, T-score $<$ 0.0001), which is shown in detail in Table~\ref{Result Table}  


\section{Discussion}
\label{Section 4}

This work focuses on continuously decoding EEG signals to predict various movement parameters during a grasp-and-lift task. To achieve this, we utilise the ERDS feature set from the EEGForceMap framework~\cite{11} and introduce an attention-based transformer model for EEG decoding. We then compare the performance of our model with that of commonly used traditional and deep learning regression models in a pseudo-online environment. Additionally, we examine how representations of movement parameters within the ERDS feature sets affect the model's overall performance.

\textit{ERDS Representation in Movement Parameters:} In this work, the EEGForceMap approach described in Section~\ref{Section 2} was developed to isolate and enhance the signal from the parietal-premotor network, which encodes neural mechanisms underlying movement planning~\cite{11}. Since prior applications of this pipeline focused on single-parameter prediction, it was initially assumed that movement parameters are encoded independently within this network. However, the ERDS representation of movement parameters has shown moderate-to-high correlations during movement encoding, as discussed in Section~\ref{Section 3}. Indicating that movement encoding is not performed separately for each parameter, but is correlated with multiple movement parameters encoded simultaneously. This might also explain PLS's reduced performance from a higher CoD of 0.425 in a prior study~\cite{11} to merely 0.13 in the current multi-parameter decoding setup (See Table~\ref{Result Table}). This drastic decline in performance can be attributed to movement parameters whose neural representations were moderately to highly correlated, as shown by ERDS. Furthermore, the correlations among the features have also impacted the performance of the MLP and TBR methods.

\textit{Comparison of Regressor Mechanisms:} The analysis revealed that the PLS regressor significantly underperformed in decoding autocorrelated parameters due to its simple architecture. In contrast, the MLP and TBR regressors performed better owing to their greater complexity, although both exhibited irregular patterns that warrant further investigation. As discussed in Section~\ref{Section 3}, the MLP-based regressor maintains consistency but at the expense of lower accuracy than in the accuracy observed in the prior study~\cite{11} while decoding a single parameter, where it achieved a CoD of 0.815 while decoding a single kinetic parameter~\cite{11}. The peculiar performance observed in the MLP regressor can be explained by the autocorrelation of the movement parameters discussed in the prior subsection, as the MLP regressor is known to examine EEG patterns independent of the temporal context~\cite{21}. Conversely, TBR leverages prior temporal patterns~\cite{21}, achieving superior performance by simultaneously decoding all movement parameters by identifying the underlying autocorrelations among them. However, this accuracy diminishes when predicting parameters separately, as these correlations are not properly captured. In conclusion, the TBR model can simultaneously assess and predict multiple movement parameters with higher accuracy and should be used in future studies. On the other hand, deep learning-based regressor models derived from the MLP regressor can be further developed to predict individual movement parameters with high accuracy.

\textit{Limitations and Future Work:} The regressor models developed in this work have demonstrated high accuracy in simultaneously decoding multiple movement parameters and are feasible for deployment, but three critical limitations remain to be addressed. First, this work explores the possibility of simultaneously decoding multiple movement parameters using different regressors. Still, the performance of these regressors needs to be validated with additional EEG data, either from another dataset. Second, recent studies have shown that those primarily influence movement parameters decoded during the grasp-and-lift task in the reaching task~\cite{13}. Future studies need to develop models that will decode these parameters, as well as the parameters required for grasping, as decoded in this work. Finally, the regressor models discussed need to be implemented in a real-world setting to check their usability. Despite these limitations, this work has demonstrated the feasibility of simultaneously decoding multiple movement parameters. 


\section{Conclusion}

This work demonstrates the feasibility of simultaneously decoding both kinetic and kinematic movement parameters from non-invasive EEG signals by integrating existing preprocessing and feature-extraction pipelines with advanced regression models. The MLP, a deep-learning-based regressor and TBR, an attention-based regressor, have been shown to decode kinetic and kinematic movement parameters simultaneously. Specifically, the TBR has shown higher accuracy by utilising prior temporal patterns to identify inherent autocorrelations. However, its inconsistent performance when decoding parameters in isolation suggests a heavy reliance on inter-parameter correlations. Conversely, the MLP offered a more stable but less accurate alternative by examining EEG patterns independently of temporal context.  Despite limited parameter diversity, our findings suggest that the TBR's high precision and low latency make it the preferred model for developing BMI applications with more intuitive and exhaustive control. The source code and dataset details are available in the EEG2GAL repository~\footnote{\url{https://github.com/HAIx-Lab/EEG2GAL}} to ensure reproducibility.


\noindent{Acknowledgement:} This study was supported by the IIT Gandhinagar startup grant (IP/IITGN/CSE/YM/2324/05). 









\bibliographystyle{IEEEtran}
\bibliography{References}

\end{document}